\documentclass[preprint,tightenlines,aps,prc,superscriptaddress,nofootinbib,preprintnumbers,showkeys,11pt]{revtex4}

\usepackage[utf8]{inputenc}
\usepackage[english]{babel}
\usepackage{graphicx}
\usepackage{physics}
\usepackage{amsmath}
\usepackage{amssymb}
\usepackage{bbm}
\usepackage{hyperref}
\usepackage{tikz}
\usepackage{xcolor}
\newcommand{\K}[1]{\ensuremath{\left(#1\right)}}
\newcommand{\Ke}[1]{\ensuremath{\left[#1\right]}}
\newcommand{\M}{\ensuremath{M_\Lambda}}
\usepackage{soul}
\newcommand{\hcancel}[1]{%
	\tikz[baseline=(tocancel.base)]{
		\node[inner sep=0pt,outer sep=0pt] (tocancel) {#1};
		\draw[black] (tocancel.south west) -- (tocancel.north east);
	}%
}%
\allowdisplaybreaks
\makeatletter\AtBeginDocument{\let\@elt\relax}\makeatother

\DeclareUnicodeCharacter{2212}{-}

\begin{document}

\title{Lifetime of the hypertriton}
\author{F. Hildenbrand}
\email{hildenbrand@theorie.ikp.physik.tu-darmstadt.de}
\affiliation{Technische Universit\"at Darmstadt, Department of Physics,
64289 Darmstadt, Germany}

\author{H.-W. Hammer}
\email{Hans-Werner.Hammer@physik.tu-darmstadt.de}
\affiliation{Technische Universit\"at Darmstadt, Department of Physics,
64289 Darmstadt, Germany}
\affiliation{ExtreMe Matter Institute EMMI, GSI Helmholtzzentrum
für Schwerionenforschung GmbH, 64291 Darmstadt, Germany}

\date{\today}

\begin{abstract}
  We calculate the lifetime of the hypertriton
  as function of the $\Lambda$ separation energy $B_\Lambda$ in an effective
  field theory with $\Lambda$ and deuteron degrees of freedom.
  We also consider the impact of new measurements of the weak decay
  parameter of the $\Lambda$.
  While the sensitivity of the total width to $B_\Lambda$ is small,
  the partial widths for decays into individual final states and the 
  experimentally measured ratio
  $R=\Gamma_{{^3\text{He}}}/\K{\Gamma_{^3\text{He}}+\Gamma_{pd}}$
  show a strong dependence. 
  For the standard value $B_\Lambda=(0.13\pm 0.05)$ MeV, we find
  $R=0.37\pm 0.05$, which is in
  good agreement with past experimental studies and theoretical calculations.
  For the recent STAR value $B_\Lambda=(0.41\pm0.12\pm0.11)$ MeV, we obtain
  $R=0.57\pm 0.11$.
\end{abstract}
\maketitle

\section{Introduction}
The addition of hyperons to nuclear bound states extends the nuclear chart into a third
dimension. These so-called hypernuclei offer a unique playground for testing our understanding low-energy
Quantum Chromodynamics in nuclei beyond the $u$ and $d$ quark sector. A particularly attractive feature of hypernuclei
is that hyperons probe the nuclear interior without being affected by the Pauli
principle. There is a vigorous experimental and theoretical program in hypernuclear physics
that dates back as far as the 1950s~\cite{Gal:2016boi}.

Here, we focus on the simplest hypernucleus, the hypertriton.
The newest results on the lifetime and binding energy of the hypertriton have created the so-called hypertriton puzzle.
The hypertriton consists of a neutron, a proton, and a $\Lambda$ particle. Its structure has been studied using hypernuclear
interaction models as well as effective field theories (See, e.g., Refs.~\cite{Congleton:1992kk,Hammer:2001ng,Wirth:2014apa,Gal:2016boi,Hildenbrand:2019sgp,Le:2019gjp}).
Furthermore, first lattice QCD calculations of light hypernuclei have become available for unphysical pion masses~\cite{Beane:2012vq}.

Since the $\Lambda$ separation energy of the hypertriton, $B_\Lambda$,
is small compared to the binding energy of the deuteron, $B_d \approx 2.2$ MeV,
it can be viewed as a $\Lambda d$ bound state at low resolution.
The most frequently cited value for this separation energy is $B_\Lambda=(0.13\pm 0.05)$ MeV \cite{Juric:1973zq},
resulting in a large separation of the $\Lambda$ from the deuteron of about $10$ fm \cite{Hildenbrand:2019sgp}.
However, recent results of the STAR collaboration  indicate that $B_\Lambda$ may be a factor three larger~\cite{Adam:2019phl}.
For a discussion of possible implications of the larger value for other hypernuclei, see Ref.~\cite{Le:2019gjp}.

\begin{figure}
		\includegraphics[width=0.9\linewidth]{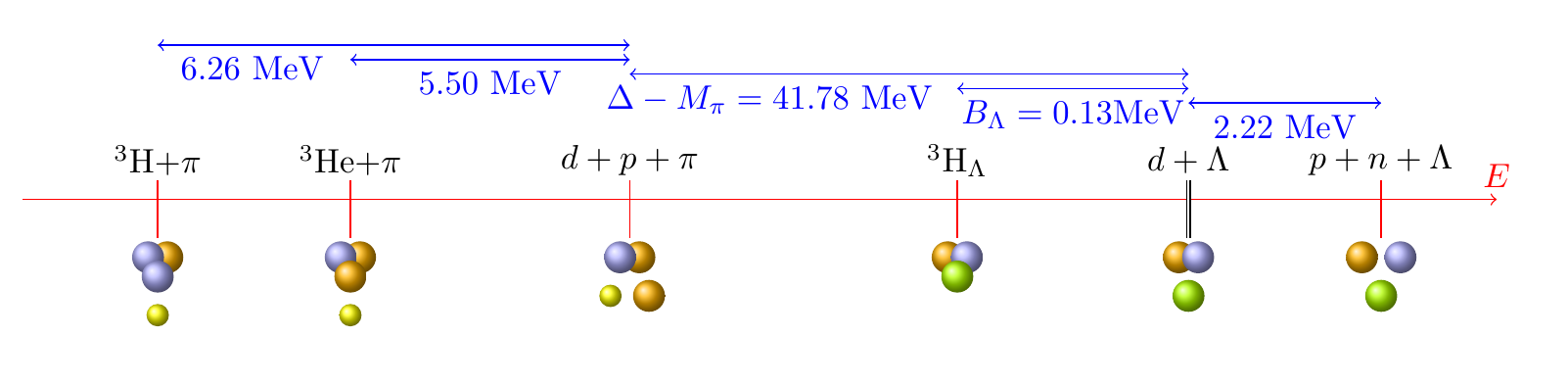}
		\caption{Most relevant thresholds for the hypertriton decay relative to the $\Lambda d$ threshold.
                  All energies are given in MeV; the figure is not up to scale.\label{fig: thresholds} }
\end{figure}

While the nucleus is stable against a breakup by strong interactions, the $\Lambda$ is unstable against weak decay
with an energy release of about $\Delta-M_\pi\approx 42$ MeV with the neutral pion mass $M_\pi=135.0$ MeV and  $\Delta=M_\Lambda-m$ the baryon mass difference, where $\M=1115.68$ MeV denotes the $\Lambda$ mass and $m=938.9$ MeV the average nucleon mass.
An overview of the most relevant thresholds is given in Fig.~\ref{fig: thresholds}.

\begin{figure}
	\begin{center}
	\includegraphics[width=0.8\linewidth]{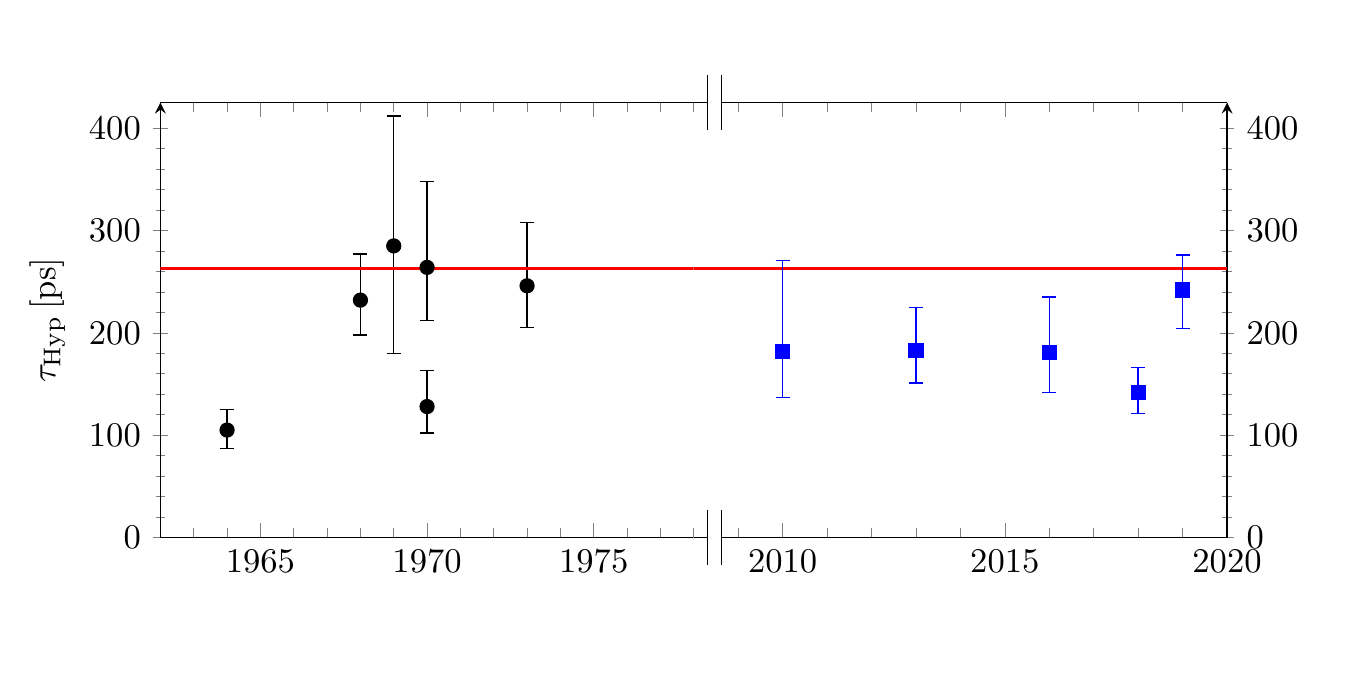}
		\caption{Compilation of lifetime measurements for the hypertriton. Blue squares show results obtained in accelerators by different collaborations \cite{Abelev:2010rv,Rappold:2013fic,Adam:2015yta,Adamczyk:2017buv,Acharya:2019qcp}. Earlier results from emulsion experiments are depicted by black circles \cite{Block:1242347,Keyes:1968zz,Phillips:1969uy,Keyes:1970ck,Bohm:1970se,Keyes:1974ev}. The red line is the PDG value for the free $\Lambda$ lifetime, $\tau_\Lambda$, as reference~\cite{Zyla:2020zbs}.\label{fig: lifetime}}
	\end{center}
\end{figure}

Experimentally, the hypertriton lifetime presents a puzzle. Old emulsion experiments give a very broad range of values
ranging from 100 ps up to 280 ps~\cite{Block:1242347,Keyes:1968zz,Phillips:1969uy,Keyes:1970ck,Bohm:1970se,Keyes:1974ev}.
Newer heavy ion experiments, tend to lie significantly below the free $\Lambda$ lifetime of about 260 ps~\cite{Abelev:2010rv,Rappold:2013fic,Adam:2015yta,Adamczyk:2017buv}. However, recent results from ALICE yield a lifetime closer to the free $\Lambda$ value~\cite{Acharya:2019qcp}.
An overview of experimental results for the hypertriton lifetime from old emulsion efforts to the newest heavy-ion experiments
is given in Fig.~\ref{fig: lifetime}.

Theoretical investigations of the hypertriton started at the same time as the first experiments~\cite{Rayet:1966fe,Ram:1971tf}.
Because $B_\Lambda$ is small compared to the deuteron binding energy, the decay of a quasifree  $\Lambda$ particle provides an
intuitive picture of the hypertriton decay and one expects that the lifetime is driven by the free $\Lambda$ width with small binding corrections.
Non-mesonic decays due to the weak $\Lambda N\to NN$ transition are strongly suppressed~\cite{Golak:1996hj,Perez-Obiol:2017nzm}.
In the $1990$s Congleton calculated the mesonic decays of the hypertriton in a $\Lambda d$  picture within the closure
approximation~\cite{Congleton:1992kk}. Assuming a most likely pion momentum, he obtained a lifetime $\tau_{{}_\Lambda^3\text{H}}$
about $10\%$ shorter than the free $\Lambda$ lifetime $\tau_\Lambda$. This calculation also hinted that the details of the
hypertriton wave function do not seem to be important. Later complete three-body Faddeev calculations using realistic hyperon-nucleon potentials
found a  $3\%$ decrease relative to $\tau_\Lambda$~\cite{Kamada:1997rv}.
Newer approaches combine the assets of both calculations, finding the impact of pionic final state interactions to be about
$10\%$ of $\Gamma_\Lambda$~\cite{Gal:2018bvq}. Recently P\'erez-Obiol et al.~calculated the channel
${}_\Lambda^3\text{H} \mapsto\pi^-+{^3\text{He}}$ based on NCSM wave functions for $^3$He and the hypertriton~\cite{Perez-Obiol:2020qjy}.
Using the experimental branching ratio $\Gamma_{^3\text{He}}/\K{\Gamma_{^3\text{He}}+\Gamma_{pd}}$~\cite{Keyes:1968zz,Keyes:1970ck,Keyes:1974ev,Block:1963} and varying $B_\Lambda$ by changing the short-distance cutoff in the NCSM,
they found that all recent experimental measurements of $B_\Lambda$ and $\tau_\Lambda$ are internally consistent within their uncertainties.

In this work, we address the hypertriton lifetime puzzle in a pionless effective field theory (EFT) approach with $\Lambda d$ degrees
of freedom. The pionless EFT framework provides a controlled, model-independent description
of weakly-bound nuclei based on an expansion in the ratio of short- and
long-distance scales (see Refs.~\cite{Beane:2000fx,Bedaque:2002mn,Epelbaum:2008ga,Hammer:2017tjm,Hammer:2019poc} for reviews).
Since leptonic decays are strongly suppressed, we focus on the $\pi$-mesonic decays of the hypertriton into nucleon-deuteron and
trinucleon final states.
This process can be treated in pionless EFT since the pions are on (or near) their mass shell \cite{Beane:2002aw}.
The choice of  $\Lambda d$ degrees
of freedom is well motivated by the separation of scales between $B_\Lambda$ and the deuteron binding energy, as
well as explicit three-body calculations in pionless EFT with $\Lambda pn$ degrees of freedom~\cite{Hammer:2001ng,Hildenbrand:2019sgp}.
Our approach has the advantage that $B_\Lambda$ enters as a free-parameter in the EFT and can be varied without changing other
observables. In particular, we investigate the properties of the hypertriton decay for $\Lambda$ separation energies
in the range $0 \leq B_\Lambda \leq 1.5$ MeV.
Furthermore, we investigate the sensitivity to  new results for
the asymmetry parameter
$\alpha_-$~\cite{Ablikim:2018zay,Ireland:2019uja} correcting the previous value by about 15\%. This quantity encodes information
on the relative contributions of parity conserving and violating parts of the interaction. Preliminary results of our work were presented in Ref.~\cite{Hammer2020}.

The structure of the paper is as follows. We start with an overview of the formalism and our procedure to
fix the low-energy constants in Sec.~\ref{sec: Form}. After that, we discuss the calculation of the two most prominent channels
for mesonic decays, a weak decay of the bound $\Lambda$ followed by the break up into a nucleon and a deuteron in Sec.~\ref{sec: Nd} and a
weak decay of the bound $\Lambda$ with a trinucleon in the final state in Sec.~\ref{sec: he}.
We follow up with a discussion of our results for the dependence of the lifetime on $B_\Lambda$ and $\alpha_-$ in Sec.~\ref{sec: res}.
We then conclude with a summary and outlook in Sec.~\ref{sec: sum}. A few calculational details are given in the Appendix.

\section{Formalism}
\label{sec: Form}
\subsection{Preliminaries}
Since the $\Lambda$ separation energy of the hypertriton,
$B_\Lambda \approx 0.13$ MeV, is small compared to the binding energy of the
deuteron, $B_d \approx 2.2$ MeV, the hypertriton can to good accuracy
be described as two-body bound state of a deuteron and a $\Lambda$ particle.
The typical momentum scale for the hypertriton can be estimated from the
energy required for breakup into a $\Lambda$ and a deuteron as
$\gamma_3^\Lambda\sim2\sqrt{\K{mB^3_\Lambda-\gamma^2_d}/3}
=2\sqrt{mB_\Lambda/3}\approx0.3\gamma_d$ with $\gamma_d=45.68$ MeV
the deuteron binding momentum and $m$ the nucleon mass. 
The breakdown scale is determined by the deuteron
breakup, which is the lowest energy process not explicitly included
in our EFT. A full three-body calculation of $\Lambda d$ scattering
in the framework of Ref.~\cite{Hildenbrand:2019sgp}, as well as
previous calculations \cite{GhaffaryKashef:1971tv}, show that
$\Lambda d$ scattering remains essentially elastic up to twice the
deuteron binding momentum $\gamma_d$.
As a consequence, the $\Lambda d$ picture can safely be used for elastic
processes up to momenta of order $2\gamma_d$, which we take as the breakdown
scale. Thus we estimate the uncertainty of our calculation as
$\sqrt{mB_\Lambda/3}/\gamma_d$.\footnote{This estimate is more
conservative than an estimate based on the $\Lambda d$ effective range
$r_{\Lambda d}\approx 1.3$~fm.}
As a consequence, the expansion parameter is 0.15 for
$B_\Lambda =0.13$~MeV, 0.25 for $B_\Lambda =0.41$~MeV, and reaches
0.5 for $B_\Lambda =1.5$~MeV.
As $B_\Lambda$ approaches the deuteron binding energy, our framework breaks down.
The $\Lambda d$ picture for the low-energy structure of the hypertriton
is also  supported by the work of
Congleton~\cite{Congleton:1992kk} and our recent investigation of the
hypertriton structure and matter radii~\cite{Hildenbrand:2019sgp},
where a three-body framework with $pn\Lambda$ and a two-body framework
with $\Lambda d$ degrees of freedom were compared.
Since the deuteron is stable, the lifetime of the
hypertriton is determined by the decay of a quasifree $\Lambda$
inside the hypertriton with small binding corrections. As discussed above,
some measurements find the lifetime of the hypertriton to be
about 30\% shorter than the lifetime of the free $\Lambda$.
The pionless EFT description of the hypertriton in the
$\Lambda d$ picture provides an appropriate starting 
point to resolve this question.

The main decay channels of the hypertriton are driven by the hadronic
decay of the $\Lambda$:
\begin{equation}\label{eq:decays}
\begin{aligned}
{}_\Lambda^3\text{H}&\mapsto\pi^-+{^3\text{He}}, & {}_\Lambda^3\text{H}&\mapsto\pi^0+{^3\text{H}},\\
{}_\Lambda^3\text{H}&\mapsto\pi^-+d+p,&{}_\Lambda^3\text{H}&\mapsto\pi^0+d+n,\\
{}_\Lambda^3\text{H}&\mapsto\pi^-+p+n+p,&{}_\Lambda^3\text{H}&\mapsto\pi^0+p+n+n.
\end{aligned}
\end{equation}
In the first line of Eq. \eqref{eq:decays}, no breakup of the three-body
nucleus takes place. Going down from top to bottom, more and more subsystems are
broken up.
The deuteron breakup processes in the third line have only a small available phase space 
and are suppressed compared to the other ones. The corresponding partial widths
are a factor 100 smaller than the other hadronic decay channels~\cite{Kamada:1997rv}.
Moreover, the non-mesonic decay branch of the hypertriton due to the reaction $\Lambda N \to NN$ is small and makes up only $1.5\%$ of
the total decay rate \cite{Golak:1996hj,Perez-Obiol:2017nzm}. It is not included in our calculation but
experimentally these decay rates cannot be separated.
Finally, note that the charged channels ($\pi^-$ in the final state) and the neutral channels ($\pi^0$ in the final state) are connected via the empirical
$\Delta I=\frac{1}{2}$ rule, setting the ratio of the channels in Eq. \eqref{eq:decays} line for line approximately equal to $2$. 

In the following, we describe the hypertriton in leading order pionless EFT with $\Lambda d$ degrees of freedom.
The typical momentum of the deuteron and the $\Lambda$ in the hypertriton,
$\gamma_3^\Lambda\approx 14$ MeV,  is small compared to the pion mass
and the deuteron binding momentum~\cite{Hildenbrand:2019sgp}.
We focus on the dominant trinucleon and nucleon-deuteron
  final states in the first two lines of Eq.~(\ref{eq:decays}),
  including the the deuteron-nucleon final state interaction in the
  $S=\frac{1}{2}$ and $S=\frac{3}{2}$ channels.
  The deuteron breakup processes in the third are unlikely to resolve the lifetime puzzle bcause of their
small branching ratio.
The pion in the outgoing state is included with relativistic kinematics due to the large energy of $\M-m-M_\pi\approx 0.26 M_\pi$ released at the weak vertex.
The final state interaction of the pion is neglected.

\subsection{Fixing the weak interaction}
\label{sec:weak}
We use the  free $\Lambda$ decay to fix the weak interaction vertex. The non-leptonic decay matrix element can be written as \cite{Holstein:1985pe}
\begin{align}
\mathcal{M}_{\Lambda\mapsto n\pi^0}=(-1)iG_FM_\pi^2\bar{u}\K{\boldsymbol{p}'}\Ke{\tilde{A}_\pi+\tilde{B}_\pi\gamma_5}u\K{\boldsymbol{p}},
\end{align} 
where $\tilde{A}_\pi$ is the parity violating (PV) amplitude while $\tilde{B}_\pi$ is parity conserving (PC).
(Note that the pion has negative parity.)
The prefactor $(-1)$ is the isospin factor for the $p\pi^0$ channel. The respective factor for the
$n\pi^-$ channel is $\sqrt{2}$. The Fermi constant is taken as $G_F=1.1664\times10^{-5}\text{GeV}^{-2}$~\cite{Tanabashi:2018oca}.
In the following, we will only calculate the width for the neutral pion channel and obtain the
width for the corresponding charged pion channel by applying isospin symmetry
and the $\Delta I=\frac{1}{2}$ rule. (See Sec.~\ref{sec: res} for more details.)
Due to the small binding momentum of the hypertriton,
it is sufficient to treat the baryons non-relativistically. The non-relativistic reduction of
the decay  matrix element is
\begin{align}
  \label{eq:weak-c}
\mathcal{W}_k\equiv\mathcal{M}_{\Lambda\mapsto n\pi^0}^{\text{reduced}}=-iG_FM_\pi^2\K{A_\pi+\frac{B_\pi}{\M+m}\boldsymbol{\sigma}\cdot\boldsymbol{k}},
\end{align}
with $\boldsymbol{k}$ the momentum of the pion and $\boldsymbol{\sigma}$ the usual Pauli spin-matrices
(see also Refs.~\cite{Kamada:1997rv,Golak:1996hj}). Note that we have redefined the amplitudes for the PC ($B_\pi$) and PV ($A_\pi$) part to absorb some normalization factors of the matrix element.

\begin{figure}[htb]
	\begin{center}
		\includegraphics*[width=0.4\linewidth]{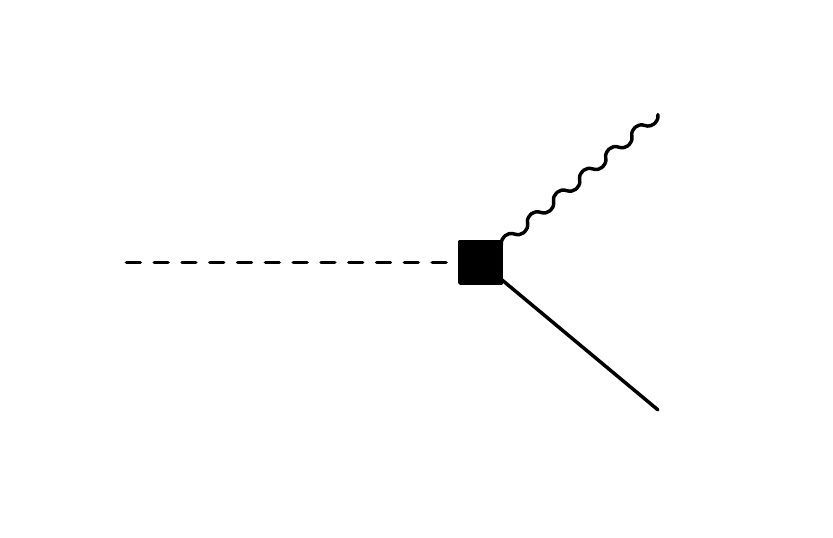}
		\caption{Vertex for the decay of the free
                  $\Lambda$ (dashed line) into a nucleon (solid line) and
                  a pion (wiggly line). The weak decay operator $\mathcal{W}_k$
                  is indicated by the black box\label{Dia: free }.}
	\end{center}
\end{figure}

\begin{figure}[htb]
	\begin{center}
\includegraphics[width=0.8\linewidth]{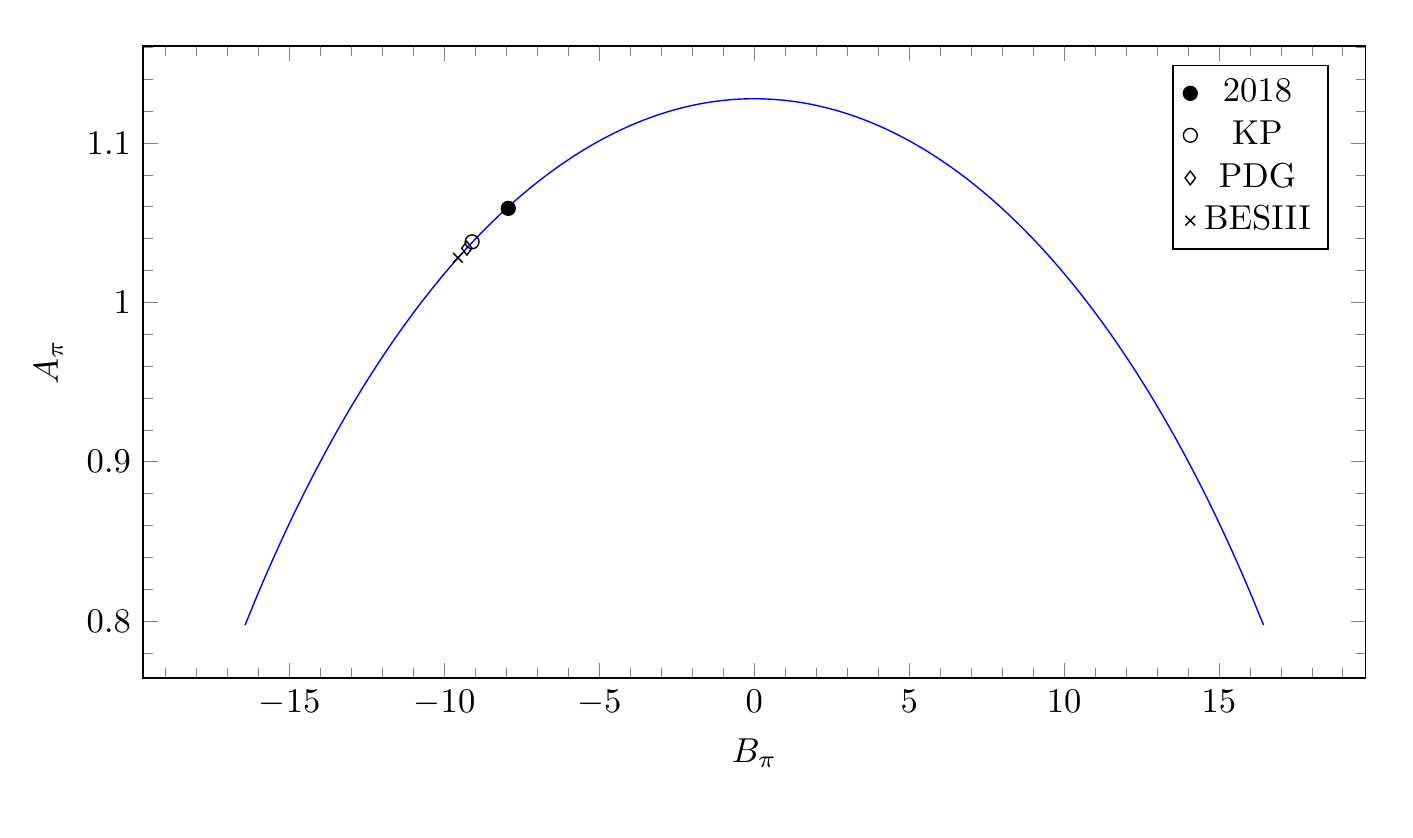}
\caption{Weak $\Lambda$-decay amplitudes $A_\pi$ and $B_\pi$
  determined by Eqs.~\eqref{eq:freewidth}, \eqref{eq:polarisation}
  as a parameter plot of the $\Lambda$-decay parameter in the range
  $-1\leq\alpha_-\leq1$. The
  free $\Lambda$ lifetime is fixed to the experimental value. Different
  experimental results for $\alpha_-$ are marked as indicated in the legend.
  \label{fig: ABalpha} }
	\end{center}
\end{figure}

It is now straightforward to calculate free width of the $\Lambda$, according to the diagram given in Fig.~\ref{Dia: free }. The $\Lambda$ is assumed to be at rest, while the momentum of the outgoing pion is denoted $\boldsymbol{k}$ and the one of the nucleon is $\boldsymbol{p}$.
We obtain:
\begin{align}
	\Gamma_\Lambda^n=\int\frac{\dd[3]{k}}{\K{2\pi}^3}\frac{\dd[3]{p}}{\K{2\pi}^3}\frac{1}{2\omega_k}\K{2\pi}^4\delta^{\K{3}}\K{\boldsymbol{p}+\boldsymbol{k}}\delta\K{\Delta-\omega_k-\frac{p^2}{2m}}\frac{1}{2}\sum_{\text{spins}}\abs{\mathcal{W}_k}^2,
\end{align}
with $\omega_k=\sqrt{M_\pi^2+k^2}$ the relativistic energy of the pion and $\Delta=\M-m$ the baryon mass difference, which is released at the weak vertex $\mathcal{W}_k$. The $\delta$ functions fix the momentum of the outgoing pion to be 
\begin{align}
\bar{k}=\sqrt{2}\sqrt{-\sqrt{m^2\K{m^2+2\Delta m+M_\pi^2}}+m^2+\Delta m}.
\end{align} 
The resulting width is then given by 
\begin{align}\label{eq:freewidth}
	\Gamma_\Lambda^n=\frac{G_F^2M_\pi^4}{2\pi}\frac{m\bar{k}}{m+\omega_{\bar{k}}} \K{A_\pi^2+\K{\frac{B_\pi}{\M+m}}^2\bar{k}^2}.
\end{align}
The associated lifetime $1/\Gamma_\Lambda=\tau_\Lambda=(263\pm2)$ ps is experimentally established very well \cite{Tanabashi:2018oca}. We use this observable together with the polarization of the $\Lambda$
\begin{align}\label{eq:polarisation}
P_\Lambda=\frac{\frac{A_\pi B_\pi}{\M+m}\bar{k}}{A_\pi^2+\K{\frac{B_\pi}{\M+m}}^2\bar{k}^2}=\frac{\alpha_-}{2},
\end{align}
which determines the $\Lambda$ decay parameter $\alpha_-$ to fix the weak  interaction strength. Up to 2018 the widely accepted value was $\alpha_-^{2018}=0.642\pm0.013$ \cite{Tanabashi:2018oca}, but new results from the BESIII Collaboration suggest a significantly higher value $\alpha_-^{\text{BESIII}}=0.750\pm0.009\pm0.004$ \cite{Ablikim:2018zay}. Also an independent estimation from kaon-photo production suggests a value of $\alpha_-^{\text{KP}}=0.721\pm0.006\pm0.005$~\cite{Ireland:2019uja} close to the results of BESIII. The two latter ones are used for the current PDG value of $\alpha_-^{PDG}=0.732\pm0.014$~\cite{Zyla:2020zbs}. The results for the PV and PC amplitudes $A_\pi$ and $B_\pi$ as
  determined by Eqs.~\eqref{eq:freewidth},~\eqref{eq:polarisation} and the
  experimental $\Lambda$ lifetime are depicted in Fig. \ref{fig: ABalpha}.
  The different
  values for $\alpha_-$ are marked explicitly.

\subsection{Hypertriton as two-body system}
The typical momentum scales of the  deuteron and the $\Lambda$ in the hypertriton are small compared to the rest masses
(see, e.g., Refs.~\cite{Hildenbrand:2019sgp,Hammer:2001ng}), so they can be treated non-relativistically. Hence single particle propagators are given by
\begin{align}
iS_{d,\Lambda,N}\K{p_0,\boldsymbol{p}}=\frac{i}{p_0-\frac{\boldsymbol{p}^2}{2M_i}+i\epsilon},
\end{align} 
with $M_i$ the respective particle masses of the deuteron and the $\Lambda$ and $M_N\equiv m$ the nucleon mass.

\begin{figure}[htb]
	\begin{center}
		\includegraphics[width=0.8\linewidth]{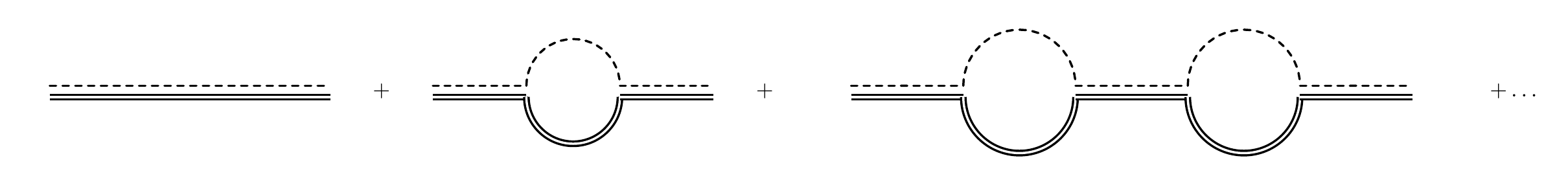}
		\caption{Diagrams contributing to the hypertriton propagator in the effective field theory
                  with deuteron (double line) and the $\Lambda$ (dashed line)
                  degrees of freedom.
                  \label{dia: geoseries} }
	\end{center}
\end{figure}
The full propagator of the interacting $\Lambda d$ system
in the dimer picture (cf.~\cite{Braaten:2004rn}) is
depicted in Fig.~\ref{dia: geoseries}. Evaluating the geometric series,
we obtain the full "dimer" propagator:
\begin{equation}
iD_{{}_\Lambda^3\text{H}}\K{p_0,\boldsymbol{p}}=\frac{2\pi}{\mu_{\Lambda d}\,g^2}\frac{-i}{-\gamma_{\Lambda d}+\sqrt{-2\mu_{\Lambda d}\K{p_0-\frac{\boldsymbol{p}^2}{2M_\Lambda+M_d}+i\epsilon}}}\,,
\end{equation}
which has a pole at the $\Lambda$ separation energy of the hypertriton,
$B_\Lambda$.
The residue of the pole is the wave function renormalization
$Z_{^3_\Lambda \text{H}}(B_\Lambda)=\frac{2\pi}{(\mu_{\Lambda d}\,g)^2}\sqrt{2\mu_{\Lambda d}B_\Lambda}$.
For convenience, we will use the reduced wave function renormalization $\bar{Z}_{^3_\Lambda \text{H}}=g^2Z_{^3_\Lambda \text{H}}$, where the coupling constant $g$ has been divided out,  in the following sections.

We now go on to calculate the weak decay of the hypertriton
in the $\Lambda d$ picture. 

\section{Nd channels \label{sec: Nd}}
\begin{figure}[htb]
	\begin{center}
		\includegraphics[width=0.8\linewidth]{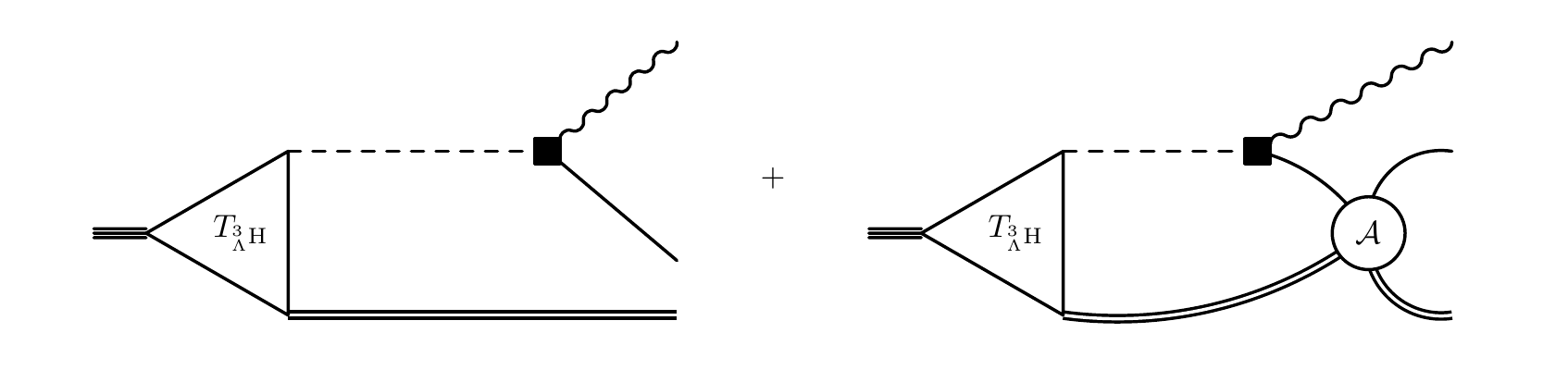}
		\caption{Decay of the hypertriton into nucleon-deuteron channels. The amplitude $\mathcal{A}$ depicts the final state interaction between the decay nucleon and the deuteron from Eq.~\eqref{eq:fsieq}.\label{dia: ndchannel} }
	\end{center}
\end{figure}

The main contribution to the hypertriton lifetime for small $\Lambda$ separation energy $B_\Lambda$ is expected to
come from the nucleon-deuteron  channels, since in the limit of vanishing $B_\Lambda$, all other channels are suppressed.
To be precise, we expect  
$\Gamma\left({}^3_\Lambda \text{H}\mapsto\right.$
\hcancel{$\pi^{-/0}+N+d$}$\big)\mapsto 0$ in the limit $B_\Lambda\mapsto0$,
because the outgoing states do not correspond to those of a free
$\Lambda$ decay plus a spectator deuteron. Therefore we need to retrieve the free $\Lambda$ width in the
limit $B_\Lambda\mapsto 0$ from the $Nd$ channels.\footnote{If the deuteron breakup is also included, decays
  into three nucleons contribute at threshold as well.} At leading order, diagrams with and without a final
state interaction between the decay nucleon and the deuteron contribute, see also Fig.~\ref{dia: ndchannel}.
We neglect pionic final state interactions, since the pions are Goldstone bosons which interact weakly.
Furthermore, all pionic scattering lengths, measured in pionic atoms or calculated in HB$\chi$PT,
are smaller than few percent of the inverse pion mass~\cite{Bernard:1995pa,Fettes:1998ud,Hauser:1998yd,Meissner:2005bz,Schwanner:1984sg,Beane:2002wk} and phase shifts are still small at the relevant
energies~\cite{Roper:1965pfb,BRAYSHAW1977139,ARVIEUX1980205}.  
Recent calculations indicate that they may change the result by up to
$10-14\%$ of the
free $\Lambda$ width \cite{Gal:2018bvq,Perez-Obiol:2020qjy}. However, this is beyond the leading order
accuracy of our calculation.
The final state interaction between the outgoing
nucleon and deuteron is described by the scattering amplitude for a shallow bound state with binding momentum
$\gamma_{Nd}$: 
\begin{align}
  \label{eq:fsieq}
\mathcal{A}\K{E}=\frac{2\pi}{\mu_{Nd}}\Ke{-\gamma_{Nd}+\sqrt{-2\mu_{Nd} E-i\epsilon}}^{-1},
\end{align}
and occurs in the $S=1/2$ and $S=3/2$ channel. In the $S=1/2$ channel  we tune  $\gamma_{Nd}$ such that the correct triton binding energy $B_{^3\text{H}}$ with respect to the $dN$
threshold is reproduced. In the $S=3/2$ channel we use the
  $nd$ scattering length $a_{3/2}=6.35$ fm~\cite{Dilg:1971gqb}. Both channels interfere with the diagram without final state interactions.
Utilizing the momentum $\delta$ function, the width $\Gamma_{{Nd}}$ is given by 
\begin{align}\label{eq:width Nd start}
\begin{split}
\Gamma_{{nd}}=&2\pi\int\int\frac{\dd[3]{p}}{\K{2\pi}^3}\frac{\dd[3]{k}}{\K{2\pi}^3}\frac{1}{2\omega_k}\frac{1}{2}\sum_{\text{spins}}\abs{\mathcal{M}\K{{}_\Lambda^3\text{H}\mapsto \pi nd}}^2\\
&\times\delta\K{\Delta-B_\Lambda-\omega_k-\frac{p^2}{2M_d}-\frac{\K{\boldsymbol{k}+\boldsymbol{p}}^2}{2m}},
\end{split}
\end{align}
with $\boldsymbol{k}$ the outgoing pion momentum and $\boldsymbol{p}$ the deuteron momentum. The invariant matrix element is the sum of the diagrams in Fig.~\ref{dia: ndchannel}: $\mathcal{M}\K{{}_\Lambda^3\text{H}\mapsto \pi Nd}=\mathcal{M}_{{Nd}}^{\text{\st{FSI}}}+\mathcal{M}_{{Nd}}^{\text{FSI}}$.
It can be most easily seen that the limit $B_\Lambda\mapsto 0$ is indeed fulfilled by neglecting the final state interaction for the moment. The matrix element $\mathcal{M}_{{nd}}^{\text{\st{FSI}}}$ is then given by 
\begin{align}
\mathcal{M}_{{nd}}^{\text{\st{FSI}}}\K{\boldsymbol{k},\boldsymbol{p}}=\sqrt{\bar{Z}_{^3_\Lambda \text{H}}(B_\Lambda)}S_\Lambda\K{-B_\Lambda-\frac{p^2}{2M_d},-\boldsymbol{p}}\mathcal{W}_k,
\end{align}
which is directly related to the normalization of the hypertriton wave function. Therefore the expression given in Eq.~\eqref{eq:width Nd start} contains a so-called Dirac series in the limit $B_\Lambda\mapsto 0$ and hence directly reduces to $\Gamma^n_\Lambda$.

Including now final state interactions and moving away from the limit $B_\Lambda\mapsto 0$ the scalar part of the matrix element $\mathcal{M}_{{nd}}^{\text{FSI}}$ reads
\begin{align}\label{Eq: marixelloop}
\begin{split}
\mathcal{M}_{{nd}}^{\text{FSI}}=&\underbrace{i\int\frac{\dd[4]{q}}{\K{2\pi}^4}S_\Lambda\K{q_0,\boldsymbol{q}}S_d\K{-B_\Lambda-q_0,-\boldsymbol{q}}S_p\K{\Delta+q_0-\omega_k,\boldsymbol{q}-\boldsymbol{k}}}_{=I_q\K{k,B_\Lambda}}\\
&\times\mathcal{A}\K{\Delta-B_\Lambda-\omega_k-\frac{k^2}{2M}}\sqrt{\bar{Z}_{^3_\Lambda \text{H}}\K{B_\Lambda}}\mathcal{W}_k.
\end{split}
\end{align}
The energy shift in the amplitude $\mathcal{A}$ is due to the boost of the
nucleon-deuteron system in the hypertriton decay. $M$ denotes here the total mass of the $nd$ system.

Now we proceed to the  evaluation of the integral $I_q\K{k,B_\Lambda}$. Due to the energy release at the weak vertex, the nucleon propagator $S_N$ has up to two poles in the $\boldsymbol{q}$  loop momentum integration depending on the angle between the outgoing pion momentum $\boldsymbol{k}$ and $\boldsymbol{q}$. We end up with the following expression
\begin{align}\label{Eq: marixelloopres}
I_q\K{k,B_\Lambda}=&\frac{2m\mu_{\text{d}\Lambda}}{k\K{2\pi}^2}\int\dd{q}q\ln\Ke{\frac{2m\mu_{\text{d}\Lambda}B_\Lambda+mq^2-2\mu_{{Nd}}qk+\frac{\mu_{{Nd}}^2}{m}k^2}{2m\mu_{\text{d}\Lambda}B_\Lambda+mq^2+2\mu_{{Nd}}qk+\frac{\mu_{{Nd}}^2}{m}k^2}}\frac{1}{q+\bar{q}}\frac{1}{q-\bar{q}}\\
&\text{with }\quad \bar{q}=\frac{1}{m}\sqrt{\mu_{{Nd}}\K{-2m^2\K{B_\Lambda+\omega_k-\Delta}+k^2\K{\mu_{{Nd}}-m}}},
\nonumber
\end{align} 
which can be evaluated utilizing the principal value method. 

The evaluation of the phase space restricts the allowed momenta since the energy delta function in Eq. \eqref{eq:width Nd start} depends on the angle between $\boldsymbol{k}$ and $\boldsymbol{p}$. Evaluating the angular integration between $\boldsymbol{k}$ and $\boldsymbol{p}$ leaves two Heaviside step functions $\Theta$ behind, restricting the area of integration. The phase space reads   
\begin{align}
\begin{split}
\rho\K{k,p}&= \frac{m k p}{\omega_k}\Ke{\Theta\K{\phi^+\K{k,p}}-\Theta\K{\phi^-\K{k,p}}}\qq{with}\\
\phi^\pm\K{k,p}&=\frac{k^2}{m}\pm \frac{2kp}{m}+\frac{p^2}{\mu_{{Nd}}}+2\K{B_\Lambda+\omega_k-\Delta}
\end{split}
\end{align}
so that 
\begin{align}
  \label{eq: pdw}
\Gamma_{{nd}}&=\frac{1}{\K{2\pi}^3}\int\int\dd{p}\dd{k}\rho\K{k,p}\frac{1}{2}\sum_{\text{spins}}\abs{\mathcal{M}\K{{}_\Lambda^3\text{H}\mapsto \pi nd}}^2.
\end{align}  
For more details see App.~\ref{app: loopdet}. We emphasize that the phase space integrals are
evaluated exactly and no closure approximation is assumed.

\section{Helium/Triton channel \label{sec: he}}

\begin{figure}[htb]
	\begin{center}
		\includegraphics[width=0.5\linewidth]{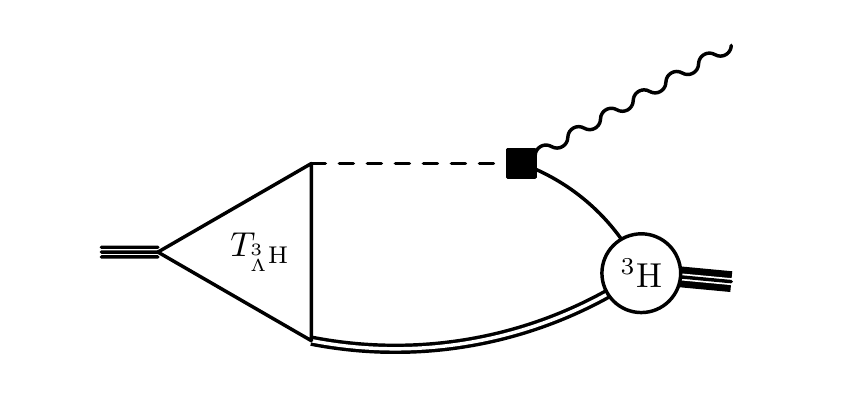}
		\caption{Decay of the hypertriton into a triton and a neutral pion.
                  A similar diagram with an outgoing helion exists in the charged decay channel.\label{dia: helium} }
	\end{center}
\end{figure}

The second contribution to the hypertriton decay in our theory
comes from decays into trinucleon final states, i.e., $^3$He and
$^3$H. As before we calculate the decay into  $^3$H and a neutral pion
and infer the charged channel using the $\Delta I=1/2$ rule.
Because the $^3$H state is on shell, only its wave function
  renormalization enters into the calculation and reproduces
  the correct asymptotic normalization constant. The details of the
  wave function do not enter.
Since we are neglecting pionic final state interactions,
there is only one diagram contributing to the width in this channel,
which is depicted in Fig.~\ref{dia: helium}.
As in the case for the free $\Lambda$, the outgoing momentum of the pion is fixed, therefore the $\Gamma_{{^3\text{H}}}$ phase space looks similar to the free one
\begin{align}
\Gamma_{{^3\text{H}}}=\int\int\frac{\dd[3]{p}}{\K{2\pi}^3}\frac{\dd[3]{k}}{\K{2\pi}^3}\frac{1}{2\omega_k}\frac{1}{2}\sum_{\text{spins}}\abs{\mathcal{M}_{{^3\text{H}}}}^2\K{2\pi}^4\delta^{\K{3}}\K{\boldsymbol{p}+\boldsymbol{k}}\delta\K{\Delta-\omega_k-\frac{p^2}{2M_{{^3\text{H}}}}}
\end{align}
 with $\Delta=M_{\text{$^3$H$_\Lambda$}}-M_{{^3\text{H}}}$ and $\boldsymbol{p}$ is now the momentum of the outgoing $^3$H nucleus. $\bar{Z}_{^3\text{H}}$ is the $^3$H wave function renormalization, constructed in a similar way to the hypertriton one. In fact we can reuse the calculation for the phase space from the free $\Lambda$ width together with the loop analysis done before for the $Nd$ case. We obtain 
 \begin{align}\label{eq: heliumw}
\Gamma_{{^3\text{H}}}=\frac{G_F^2M_\pi^4}{\pi}\frac{\bar{k} M_{{^3\text{H}}}}{M_{{^3\text{H}}}+\omega_{\bar{k}}}\bar{Z}_{^3_\Lambda \text{H}}(B_\Lambda)\bar{Z}_{{^3\text{H}}}\K{B_{{^3\text{H}}}}\K{A_\pi^2+\frac{1}{9}\K{\frac{B_\pi}{\M+m}}^2\bar{k}^2}\abs{I_q\K{\bar{k},B_\Lambda}}^2.
 \end{align}
Using relativistic kinematics, the momentum of the outgoing pion is fixed to
\begin{align}
\bar{k}=\frac{\sqrt{\K{M_{{}_\Lambda^3\text{H}}^2+M_{{^3\text{H}}}^2-M_\pi^2}^2-4M_{{}_\Lambda^3\text{H}}^2M_{{^3\text{H}}}^2}}{2M_{{}_\Lambda^3\text{H}}}.
\end{align}

\section{results\label{sec: res}}

\subsection{Partial decay width and dependence on $\alpha_-$}
\begin{table}[htb]
	\begin{tabular}{|c|c|c|c|c|}
	\hline
	$\alpha_-$&$0.642$&$0.721$&$0.732$&$0.750$\\
	\hline
	\hline
	$A_\pi$&$\phantom{-}1.05996$&$\phantom{-}1.03759$&$\phantom{-}1.03402$&$\phantom{-}1.02789$\\
	$B_\pi$&$-7.94169$&$-9.11119$&$-9.28214$&$-9.56708$\\
	
	\hline
\end{tabular}
	\caption{Values for $A_\pi$ and $B_\pi$ for different $\alpha_-$ and  $\tau_\Lambda=263.2$ ps  (see discussion in subsection \ref{sec:weak}).\label{tab:AB}}        
\end{table}
In our calculation, we use the free $\Lambda$ lifetime, $\tau_\Lambda=263.2$ ps, and
the $\Lambda$ decay parameter $\alpha_-$ to fix the values of the weak couplings $A_\pi$ and $B_\pi$ in
Eq.~\eqref{eq:weak-c}.
The corresponding couplings for different input values of $\alpha_-$
discussed in subsection \ref{sec:weak} are given in Table~\ref{tab:AB}.
The remaining momentum integrals in in the expressions for the widths, Eqs.~\eqref{eq: pdw} and  \eqref{eq: heliumw},
are evaluated numerically, exploiting the correlation between charged and uncharged decay channels
from the $\Delta I=1/2$ rule to obtain the full rate.

\begin{figure}[htb]
	\begin{center}
		\includegraphics[width=\linewidth]{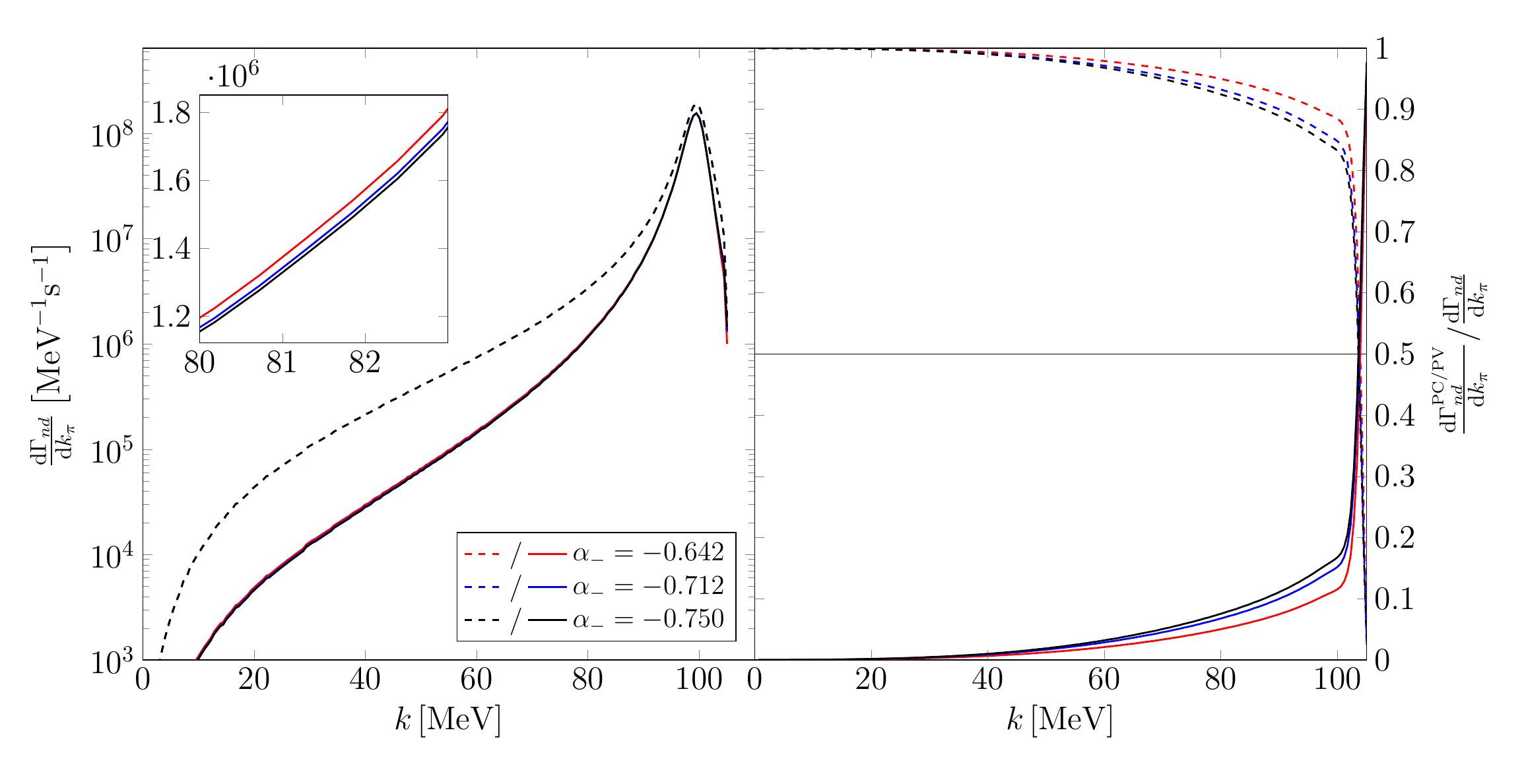}
		\caption{Left panel: logarithmic plot of the differential rate
                  $\dv{\Gamma_{{nd}}}{k}$ for different values of $\alpha_-$ indicated in the legend at fixed
                  $\Lambda$ separation energy, $B_\Lambda=0.13$ MeV. Results including (excluding) final state interactions are
                    shown by solid (dashed) lines, respectively. The inset shows the small dependence on $\alpha_-$. Right panel: relative contribution of the parity conserving (solid lines) and  parity violating part (dashed lines)
                    to the full differential rate. \label{fig: pdw}}
	\end{center}
\end{figure}

The importance of the $Nd$ final state interaction in the $S=1/2$ as well as in the $S=3/2$ channel in the hypertriton decay 
can be visualized by plotting the differential rate $\dv{\Gamma_{{nd}}}{k}$,
where $k$ is the final pion momentum for fixed $B_\Lambda$. The result for $B_\Lambda=0.13$ MeV is depicted in the left panel of Fig.~\ref{fig: pdw}.
For small pion momenta the $Nd$ final state interactions (solid lines) reduce the differential width by an order of
magnitude compared to the calculation without  final state interactions (dashed lines).
The new larger $\Lambda$ decay parameter $\alpha_-$ shifts the partial widths slightly upwards as shown in
the inset of Fig.~\ref{fig: pdw}, but the overall sensitivity is small.
It is instructive to consider the parity conserving and violating parts separately. 
Indeed, the $15\%$ change in the decay parameter shifts the contribution of the parity violating part moderately
for high $k$, as indicated in the right panel of Fig.~\ref{fig: pdw}. The parity conserving part gives a smaller
contribution over the full range of pion momenta $k$ but shows roughly the opposite behavior.
Hence, although the relative contribution of the parity violating term and the parity conserving term change moderately,
their sum only changes slightly as seen in the left panel of Fig.~\ref{fig: pdw}. This behavior is expected from the scaling behavior
of  Eqs.~\eqref{eq: pdw}, \eqref{eq: heliumw} with $A_\pi$ and $B_\pi$,
where $B_\pi$ directly scales with $k^2$. A similar trend is reflected in the partial widths
discussed below.

\subsection{Width results and comparison with theory and experiment}
\begin{figure}[htb]
	\begin{center}
		\includegraphics[width=\linewidth]{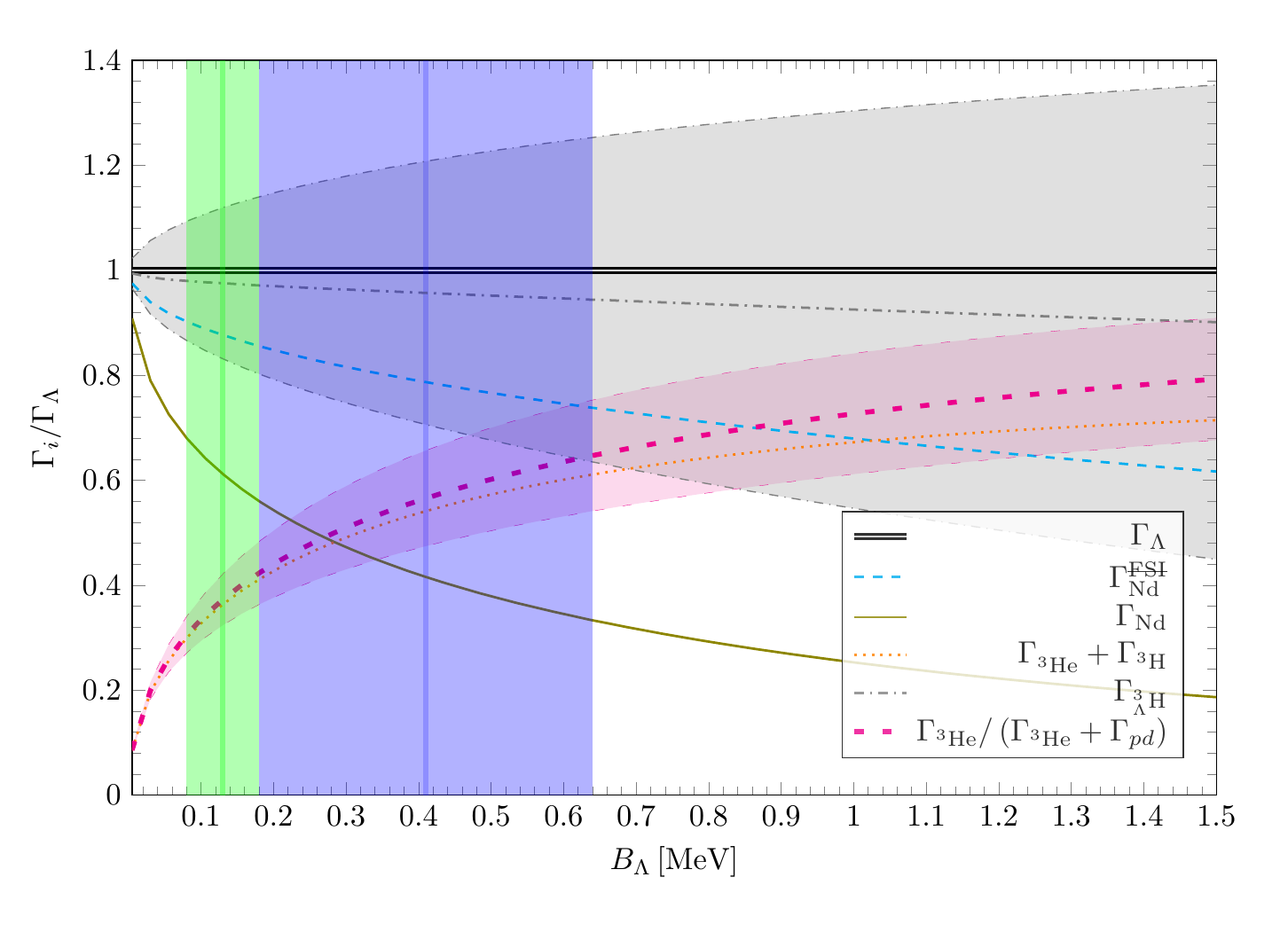}
		\caption{Partial decay widths $\Gamma_{i}$ for $\alpha_-^{PDG}$ in units of the free $\Lambda$ width $\Gamma_\Lambda$
                  as a function of the $\Lambda$ separation  energy $B_\Lambda$.
                  The ratio $R=\Gamma_{^3\text{He}}/\K{\Gamma_{^3\text{He}}+\Gamma_{pd}}$
                  is also shown. The experimental values
                  $B_\Lambda=(0.13\pm0.05)$ MeV \cite{Juric:1973zq} and $B_\Lambda=(0.41\pm0.12\pm0.11)$ MeV \cite{Adam:2019phl}
                  are indicated by the shaded light (green) and dark (blue) rectangular areas, respectively.
                  $\Gamma_{\text{Nd}}^{\text{\st{FSI}}}$ gives
                    the partial width into Nd in the absence of final state interactions.
                    For clarity, the EFT uncertainties are shown by bands only for $R$ and $\Gamma_{\text{$^3$H$_\Lambda$}}$.}
\label{fig: Widths}.
	\end{center}
\end{figure}

The results for the different partial widths are summarized in Fig.~\ref{fig: Widths}. The two prominent experimental values for
the $\Lambda$ separation energy,
$B_\Lambda=(0.13\pm0.05)$ MeV \cite{Juric:1973zq} and $B_\Lambda=(0.41\pm0.12\pm0.11)$ MeV \cite{Adam:2019phl} are
indicated by the shaded light (green)
and dark (blue) rectangular areas, respectively.
The calculated partial widths and ratios  are explained in the legend. 
To avoid cluttering the figure, the uncertainties from the effective theory expansion are shown
  only for the total width $\Gamma_{{}_\Lambda^3\text{H}}$
  and the branching ratio $R=\Gamma_{^3\text{He}}/\K{\Gamma_{^3\text{He}}+\Gamma_{pd}}$. For all partial widths the error is given by
  $\sqrt{mB_\Lambda/3}/\gamma_d$ as discussed in Sec.~\ref{sec: Form},  which corresponds to
  15\% for $B_\Lambda = 0.13$ MeV.
For very small $B_\Lambda$ the $Nd$ channel dominates, since the allowed phase space for the decay into a bound state
is smaller and for
$B_\Lambda\mapsto0$ the decay into a trinucleon state is suppressed. As $B_\Lambda$ increases, the decay  into a trinucleon
bound state
becomes more and more dominant.
While the limit $B_\Lambda \mapsto\Delta$ is outside the range of applicability of an
effective theory with $\Lambda d$ degrees of freedom, both partial decay 
widths go to zero for $B_\Lambda \mapsto\Delta$ as expected from phase space 
considerations. In this limit the hypertriton becomes stable against the 
weak decay, since the energy release at the weak vertex would be below 
the $\Lambda$ separation energy.

While the full hypertriton width $\Gamma_{\text{$^3$H$_\Lambda$}}$ does only moderately depend on $B_\Lambda$, and the correlation
appears small, the partial widths show a strong dependence. As a consequence,  the experimentally measured ratio
of the partial width into $^3$He divided by the partial width into  $^3$He and $pd$,
$\Gamma_{^3\text{He}}/\K{\Gamma_{^3\text{He}}+\Gamma_{pd}}$, is also very sensitive to $B_\Lambda$.
Hence this quantity appears to be better suited to
determine $B_\Lambda$ indirectly than the total width~\cite{Keyes:1974ev,Congleton:1992kk}.

The partial widths for the $\Lambda$ separation energies $B_\Lambda=0.13$ MeV and $B_\Lambda=0.41$ MeV
for the old and new PDG values of $\alpha_-$ are listed in
Table~\ref{tab:lifetime}. The uncertainty of the partial widths from higher orders in the EFT expansion is given by
  $\sqrt{mB_\Lambda/3}/\gamma_d$ as discussed in Sec.~\ref{sec: Form}.
Standard error propagation leads to an absolute uncertainty of $0.05$ and $0.11$ respectively
in the ratio $R=\Gamma_{^3\text{He}}/\K{\Gamma_{^3\text{He}}+\Gamma_{pd}}$ given
in the second last line of Table~\ref{tab:lifetime}.
\begin{table}[htb]
	 \renewcommand*{\arraystretch}{1.2}
	\begin{center}
		\begin{tabular}{|c|c|c|c|c|}
	\hline
	Observable&\multicolumn{2}{c|}{$B_\Lambda=0.13$ MeV}&\multicolumn{2}{c|}{$B_\Lambda=0.41$ MeV}\\
	\hline
	$\alpha_-$&\phantom{-}$\phantom{-}0.642\phantom{-}$&$\phantom{-}0.732\phantom{-}$&$\phantom{-}0.642\phantom{-}$&$\phantom{-}0.732\phantom{-}$\\
	\hline
	\hline	
	$\K{\Gamma_{{pd}}+\Gamma_{{nd}}}/\Gamma_\Lambda$&$0.612$&$0.612$&$0.415$&$0.416$\\
	$\K{\Gamma_{{^3\text{He}}}+\Gamma_{^3\text{H}}}/\Gamma_\Lambda$& $0.382$&$0.363$&$0.569$&$0.541$\\
	$\Gamma_{{}_\Lambda^3\text{H}}/\Gamma_\Lambda$& $0.992$&$0.975$&$0.984$&$0.956$\\
	$\Gamma_{{^3\text{He}}}/\K{\Gamma_{^3\text{He}}+\Gamma_{{pd}}}$& $0.384$&$0.373$&$0.578$&$0.566$ \\
	\hline
	\hline
	$\tau_{{}_\Lambda^3\text{H}}[\text{ps}]$& $264.7$&$269.8$&$267.6$&$275.0$ \\
	\hline
\end{tabular}
	\end{center}
        \caption{Widths and lifetimes for two binding energies for different $\alpha_-$. The results assume the empirical isospin rule   \label{tab:lifetime}. The widths are given as a fraction of the $\Lambda$ free width corresponding to $\tau_\Lambda=263.2$ ps. All lifetimes are given in ps. EFT uncertainties are discussed in the main text.}
\end{table}
Our results with $\alpha_-^{2018}$ compare very well with the result obtained by
Ref.~\cite{Kamada:1997rv}. Note that the peak of the differential decay width is slightly shifted due to the different particle thresholds. Considering only the phase space it seems reasonable that the width is decreasing for larger $B_\Lambda$ since the available phase space gets smaller. The result obtained by Congleton \cite{Congleton:1992kk} is in agreement with the ratio $R$, which was measured before
\cite{Keyes:1968zz,Keyes:1970ck,Keyes:1974ev,Block:1963}. However, the total width is about $15\%$ higher. Although the decay constant changes by about $15\%$ compared to the old value $\alpha_-^{2018}$, the impact on the decay rates is much smaller for small binding energies $B_\Lambda$. While the change of the partial decay width is in the order of a few percent, the total width changes barely at all. We note that the Coulomb interaction is not included explicitly in this calculation, which might shift the lifetime in the charged channel. However, part of the Coulomb interaction is included implicitly due to the tuning of $\gamma_{Nd}$ to reproduce the correct trinucleon binding energy (see Eq.~\eqref{eq:fsieq}). Our calculation supports the picture that for small  $B_\Lambda$ the lifetime of the hypertriton is mainly determined by the free $\Lambda$  lifetime with some small corrections.

The results of this work compare differently to the recent heavy ion collision experiments. Our results for low
binding energy $B_\Lambda$ lie within the error bars of the value close to the free $\Lambda$ width \cite{Acharya:2019qcp},
while other measurements tend to lie lower \cite{Abelev:2010rv,Rappold:2013fic,Adam:2015yta,Adamczyk:2017buv}.
Despite giving values for the lifetime within a large range $60-400$ ps (see also Fig.~\ref{fig: lifetime}),
older emulsion experiments give relatively consistent experimental values for the branching ratio
$R=\Gamma_{^3\text{He}}/\K{\Gamma_{^3\text{He}}+\Gamma_{pd}}$ ranging from $R=0.30\pm0.07$ to $0.39\pm 0.07$ \cite{Keyes:1968zz,Keyes:1970ck,Keyes:1974ev,Block:1963}. Both values are in agreement with our value $R|_{B_\Lambda=0.13~{\rm MeV}}=0.37\pm0.05$ for $B_\Lambda=0.13$ MeV, while
the ratio $R|_{B_\Lambda=0.41~{\rm MeV}}=0.57\pm 0.11$ comes out much larger, see also
Table~\ref{tab:lifetime}. Further on, this value is larger than the value of $R^{\text{STAR}}=0.32\pm0.05\pm0.08$ reported by STAR~\cite{Adamczyk:2017buv}.
Requiring consistency with the experimental  $R$ values, our calculation thus favors smaller binding energies.
Taking into account the uncertainty in our calculation and the experimental errors for $R$, however,
the recent STAR result $B_\Lambda=(0.41\pm0.12\pm0.11)$ MeV~\cite{Adam:2019phl} cannot be excluded.

\subsection{Effects of isospin splitting}
A discussed above, we have explicitly calculated the charged pion channels and estimated the neutral
pion channels by applying the empirical $\Delta I=1/2$ rule.
We used an average nucleon mass, the neutral pion mass $M_{\pi^0}=135.0$ MeV and neglected the Coulomb repulsion between
the deuteron and the proton. To estimate the accuracy of this approximation, we also calculated the charged channels explicitly
using the charged pion mass and the triton binding energy as input. The latter leads to a change in the
final state trinucleon binding momentum $\gamma_{Nd}$ in Eq.~\eqref{eq:fsieq} of about $10\%$.
This change, however, is absorbed completely by kinematic changes and differences in the masses.
Overall, we obtain a shift by less than $1\%$ downwards for the sum of the channels decaying into a deuteron,
while the the width for decay into the trinucleon bound states goes up by about $2\%$.
Hence the correction to the total width is negligibly small ($<0.1\%$).
The ratio $R$ moves up slightly, resulting in $R=0.38$. This shift is
significantly smaller than the estimated uncertainty
of our leading order calculation.

\section{Summary and Outlook\label{sec: sum}}
In this work, we have investigated the dependence of the hypertriton lifetime on the $\Lambda$ separation energy with an
pionless EFT with deuteron, nucleon, and $\Lambda$ degrees of freedom. The validity of such a picture for 
the low-energy structure of the hypertriton was justified in a recent investigation of the
hypertriton structure and matter radii~\cite{Hildenbrand:2019sgp},
where a three-body framework with $pn\Lambda$ and a two-body framework
with $\Lambda d$ degrees of freedom were compared in the context of pionless EFT.
The EFT framework allows us to vary
the $\Lambda$ separation energy while keeping all other low-energy constants constant. The uncertainty in the partial widths from
higher-order contributions in the $\Lambda d$ picture
is estimated to be of order  
  15\% at $B_\Lambda = 0.13$ MeV and
  25\% at $B_\Lambda = 0.41$ MeV. It can be reduced by going beyond the leading
order in the EFT expansion.

We focus on the dominant hadronic
decay channels with $\pi Nd$ and $\pi$-trinucleon final states. These channels make up 97.4\% of the total width
of the hypertriton \cite{Kamada:1997rv} and thus provide the key to understanding the hypertriton lifetime
puzzle.  We explicitly calculate the decay channels with neutral pions in the final state, evaluating all
phase space intergrals exactly. The $\Delta I=1/2$ rule allows us to obtain the full decay rate by
relating the charged and uncharged channels. An explicit calculation
of the charged channels neglecting the Coulomb interactions in the final state indicates that the corrections to
the $\Delta I=1/2$ rule are indeed small.
Preliminary results of our work were presented in Ref.~\cite{Hammer2020}.

We find
agreement with an earlier calculation by Kamada et al. in a three-body Faddeev approach in the isospin symmetry limit using
realistic Hyperon-Nucleon potentials~\cite{Kamada:1997rv}. Moreover, the calculation of Congleton~\cite{Congleton:1992kk},
who used a $\Lambda d$ picture in the closure approximation,
agrees with ours within the EFT uncertainties.
We also investigate the impact of recent changes in the weak decay parameter $\alpha_-$,
correcting the previous value by 15\%~\cite{Ablikim:2018zay,Ireland:2019uja}. While there are moderate changes in the
parity conserving and parity violating contributions, the change in the total rate is small.

For the commonly accepted value of the $\Lambda$ separation energy,  $B_\Lambda=(0.13\pm 0.05)$~MeV~\cite{Juric:1973zq},
we find the hypertriton width $\Gamma_{{}_\Lambda^3\text{H}}=(0.975\pm0.15)~\Gamma_\Lambda$, depending on the input value
for $\alpha_-$, to be close to the  free $\Lambda$ width. Varying $B_\Lambda$ between zero and 1.5~MeV, the width
decreases, reaching 90\% of the free $\Lambda$ width at $B_\Lambda=2$~MeV. 
Due to the decreasing phase space as $B_\Lambda$ increases, it must eventually vanish
as $B_\Lambda$ approaches $\Delta -M_\pi$. For physically reasonable values of $B_\Lambda$, the 
lifetime of the hypertriton is not very sensitive to $B_\Lambda$.
However the partial widths and the experimentally measured branching ratio
$R=\Gamma_{^3\text{He}}/\K{\Gamma_{^3\text{He}}+\Gamma_{pd}}$ depend strongly on the $\Lambda$ separation energy.
Our result of $R|_{B_\Lambda=0.13~{\rm MeV}}=0.37\pm 0.05$ is consistent with the experimental measurements of
$R$~\cite{Keyes:1968zz,Keyes:1970ck,Keyes:1974ev,Block:1963,Adamczyk:2017buv}, which  favor small
$\Lambda$ separation energies. The result for $R$ at  the recent STAR
value $B_\Lambda=(0.41\pm0.12\pm0.11)$ MeV~\cite{Adam:2019phl}, $R|_{B_\Lambda=0.41~{\rm MeV}}=0.57\pm 0.11$,
comes out significantly higher. 
Moreover, this value is larger than the value of $R^{\text{STAR}}=0.32\pm0.05\pm0.08$ reported by
the STAR collaboration~\cite{Adamczyk:2017buv}.
Taking into account the experimental errors and the uncertainty from
higher orders in our calculation, we can not exclude the STAR result
$B_\Lambda=(0.41\pm0.12\pm0.11)$ MeV~\cite{Adam:2019phl} but there is
some tension.

An investigation similar in spirit to ours was carried out by P\'erez-Obiol et al.~\cite{Perez-Obiol:2020qjy}.
They calculated the width for decay into a charged pion and $^3$He using NCSM wave functions from chiral EFT interactions for the
hypertriton and the helion, including final state interactions. Using the  $\Delta I=1/2$ rule and the experimental
value for $R$ as input, they determined the full hypertriton width. Varying the  $\Lambda$ separation energy by adjusting the
ultraviolet cutoff in the NCSM calculation, they calculated the width for different values of  $B_\Lambda$. UV convergence of the two-body rates
could not be fully achieved for all considered values of
$B_\Lambda$ and they had to rely on an UV extrapolation.
In this framework, the admixed $\Sigma NN$ components in the
hypertriton wave function from  $\Lambda N \leftrightarrow \Sigma N$
conversion are important and change the purely $\Lambda NN$ value by about
10\%.\footnote{At the resolution scale of pionless EFT,
  the $\Lambda N \leftrightarrow \Sigma N$ conversion is a short-range
  process that is captured in the low-energy constants of the
  theory~\cite{Hildenbrand:2019sgp}.}
Their calculation suggests that the STAR values for 
$B_\Lambda$ and $R$ are fully consistent with each other. The slight tension between our calculation
and Ref.~\cite{Perez-Obiol:2020qjy} deserves further study, especially regarding the
different dynamical inputs and strategies in the calculations.

In the EFT calulation, this requires the inclusion of higher orders. The first correction would come from the $\Lambda d$ effective
range which can be taken from Ref.~\cite{Hildenbrand:2019sgp}. In order to calculate the contribution from the
deuteron breakup channel a four-body calculation of the hypertriton
decay with $pn\Lambda\pi$ degrees of freedom is required. According to Refs.~\cite{Gal:2018bvq,Perez-Obiol:2020qjy}
pionic final state interactions could affect the width at the 10\% level which would also be relevant at next-to-leading
order. Here it might be easier to return to a theory with a fundamental deuteron to reduce complexity.

\begin{acknowledgments}
  We thank M.~G\"obel and W.~Elkamhawy for useful discussions
  and A.~Gal for comments on the manuscript.
	This work was funded by the Deutsche Forschungsgemeinschaft (DFG,
	German Research Foundation) - Projektnummer 279384907 - SFB 1245 and
	the Federal Ministry of Education and Research (BMBF) under contracts
	05P15RDFN1 and 05P18RDFN1. Moreover, it received funding from the
        European Union’s Horizon 2020 research and innovation programme under grant agreement No 824093.
\end{acknowledgments}   

\appendix

\section{Calculation details\label{app: loopdet}}
In order to evaluate the loop integral given in Eq.~\eqref{Eq: marixelloop}, we perform the $q_0$ integration with the means of standard contour integration resulting in an integral containing two factors
\begin{align}
I_q\K{k,B_\Lambda}=&\int\frac{\dd[3]{q}}{\K{2\pi}^3}\Ke{-B_\Lambda-\frac{q^2}{2\mu_{\text{d}\Lambda}}}^{-1}\Ke{\Delta-B_\Lambda-\omega_k-\frac{q^2}{2m_d}-\frac{\K{\boldsymbol{q}-\boldsymbol{k}}^2}{2m}}^{-1}.
\end{align}
Due to the positive energy $\Delta$ and the dependence on $\boldsymbol{q}\cdot\boldsymbol{k}$ the second term has a complex pole structure with up to two poles, which can in principle fall on top of each other, depending on the angle between the loop momentum $\boldsymbol{q}$ and the external momentum of the pion $\boldsymbol{k}$. In contrast, the first term is always negative, and therefore never develops a pole. Hence it is adroit to shift the angular dependence to the first term, leading to
\begin{align}
\begin{split}
I_q\K{k,B_\Lambda}=&\int\frac{\dd[3]{q}}{\K{2\pi}^3}\Ke{-B_\Lambda-\frac{q^2}{2\mu_{\text{d}\Lambda}}-\frac{\mu_{{Nd}}\boldsymbol{q}\cdot\boldsymbol{k}}{m\mu_{\text{d}\Lambda}}-\frac{\mu_{{Nd}}^2}{2m^2\mu_{\text{d}\Lambda}}k^2}^{-1}\\
&\times\Ke{\Delta-B_\Lambda-\omega_k-\frac{q^2}{2\mu_{{Nd}}}+\frac{\mu k^2}{2m^2}-\frac{k^2}{2m}}^{-1}\,.
\end{split}
\end{align}
The  angular integration can now be done independently of the second propagator and one obtains
Eq.~\eqref{Eq: marixelloopres}.

\bibliography{lifetimelib.bib}
\end{document}